\documentclass[12pt]{article} 
\usepackage[utf8]{inputenc}
\usepackage[T1]{fontenc}
\usepackage[ngerman, english]{babel}
\usepackage{natbib}
\usepackage{amsmath,amssymb,amsthm,url}
\usepackage[english=american]{csquotes}
\usepackage{mathdots}
\usepackage{multicol}
\usepackage{graphicx}
\usepackage{enumitem}  

\usepackage{booktabs}
\usepackage{multirow}

\usepackage[linesnumbered,lined,boxed,commentsnumbered]{algorithm2e}
\usepackage{scrhack}
\usepackage{pdfpages}

\usepackage{xspace}

\usepackage{color}
\usepackage{float}

\usepackage{graphics}
\usepackage{latexsym}
\usepackage{epsf}

\usepackage[linktocpage,
            colorlinks=false,linkcolor=black,
            citecolor=black,urlcolor=black,
            pdfpagemode=UseNone,
            bookmarks=true,
            bookmarksopen=false,
            hyperfootnotes=false,
            pdfhighlight=/I,
            hidelinks
            ]{hyperref}

\theoremstyle{plain}

\theoremstyle{definition}

\theoremstyle{remark}

\newcommand{\seqnum}[1]{\href{https://oeis.org/#1}{\rm \underline{#1}}}

\allowdisplaybreaks
\def\modd#1 #2{#1\ \mbox{\rm (mod}\ #2\mbox{\rm )}}

\newcommand{\Schritte}[1]{\overset{#1}{\rightarrow}}

\begin{document}
\begin{center}
\vskip 1cm{\Large\textbf 
{Castor Ministerialis}
}
\vskip 1cm
\large
Christian Hercher\\
Institut f\"{u}r Mathematik\\
Europa-Universit\"{a}t Flensburg\\
Auf dem Campus 1c\\
24943 Flensburg\\
Germany \\
\href{mailto:christian.hercher@uni-flensburg.de}{\texttt{christian.hercher@uni-flensburg.de}} \\
\end{center}

\begin{abstract}
The famous problem of Busy Beavers can be stated as the question on how long a $n$-state Turing machine (using a 2-symbol alphabet or---in a generalization---a $m$-symbol alphabet) can run if it is started on the blank tape before it holds. Thus, not halting Turing machines are excluded. For up to four states the answer to this question is well-known. Recently, it could be verified that the widely assumed candidate for five states is in fact the champion. And there is progress in searching for good candidates with six or more states.

We investigate a variant of this problem: Additionally to the requirement that the Turing machines have to start from the blank tape we only consider such Turing machines that hold on the blank tape, too. For this variant we give definitive answers on how long such a Turing machine with up to five states can run, analyze the behavior of a six-states candidate and give some findings on the generalization of Turing-machines with $m$-symbol alphabet.
\end{abstract}

\section{Basics -- What is a Turing machine?}
In his famous lecture at the International Conference of Mathematicians 1900 in Paris, Hilbert listed 23 problems, all of them at that time unsolved. One of them he described in the following way, \cite{Hilbert}:

\begin{quote}
10. Entscheidung über die Lösbarkeit einer diophantischen Gleichung

Eine \emph{diophantische} Gleichung mit irgend welchen Unbekannten und mit ganzen rationalen Zahlencoeffizienten sei vorgelegt: man soll ein Verfahren angeben, nach welchem sich mittelst einer endlichen Anzahl von Operationen entscheiden läßt, ob die Gleichung in ganzen rationalen Zahlen lösbar ist. 
\end{quote}

An English translation by Newson can be found in \cite{Hilbert2}:

\begin{quote}
10. Determination of the Solvability of a Diophantine Equation

Given a diophantine equation with any number of unknown quantities and with rational integral numerical coefficients: \emph{To devise a process according to which it can be determined by a finite number of operations whether the equation is solvable in rational integers.} 
\end{quote}

In the years after Hilbert formulated his program for mathematics it became clear that for a solution of his 10'th problem  it has to be clarified what operations are allowed, and how one can model such a computation. In fact, several such models emerged, for example recursion theory by Gödel or lambda calculus by Church. (They are all equivalent as Turing later proved.) Now the question shifted a bit and it was asked by \cite{AckermannHilbert} if there is an algorithm that decides in a finite number of steps if a given first-order logical expression is universally valid. From the German authors the name \emph{Entscheidungsproblem} (English: \enquote{decision problem}) was coined. Church and Turing gave independent proofs that no such algorithm can exist; \cite{Turing} introduced in his work the now widely accepted model for computation, the machines named after him. (Later, in 1970, Davis, Matiyasevich, Putnam, and Robinson proved that no such algorithm solving Hilberts 10'th problem can exist.)

In the following we give a slightly more modern description of Turing machines than Turing had wrote himself, like \cite{KnuthTAOCP1, Schoning09}.

 A Turing machine consists of an infinitely long tape divided into cells, with a symbol in each cell. Furthermore, it has a read-write head (RW-head), which  in each step can read the symbol of the cell over it is positioned (and write something in it). Thirdly, the Turing machine is always in a state (which is an element of a predefined set of states) and has a state table.

\begin{figure}[h]
\begin{centering}
\includegraphics{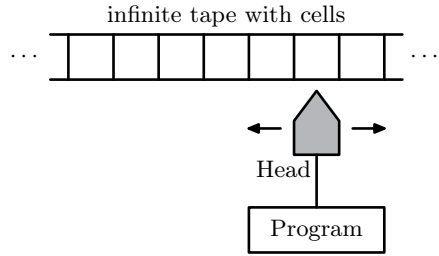}
\caption{Schematic visualization of a Turing machine}
\end{centering}
\end{figure}

When the Turing machine reads a symbol, it checks the state table to determine what (depending on the current state and the read symbol) happens next: Which symbol will be written into the current cell (which may be identical to the read symbol), will the RW-head move one cell to the left or right, and into which new state will the machine transition for the next cycle? Two states are especially important: a starting state, in which the machine begins the computation, and a halting state, which marks the end of the computation when the machine enters it. There is also a special symbol: the blank symbol. At the beginning of the computation, all cells except for a finite number must be filled with this symbol.

Turing machines represent \textsc{the} model of computability in theoretical computer science. Therefore, the question arises of what is computable and which functions, for example, are not. The Hungarian mathematician \cite{Rado} posed the question of Busy Beavers: Given the number $n$ of states (plus an additional halting state) of a Turing machine, and assuming that there is only one additional symbol besides the blank symbol and that the tape is completely empty at the start of the computation, how many steps can such a machine take before it halts? (In particular, the question requires that the computation terminates after a finite time.) For small values of $n$, this question can still be answered easily. But for $n=5$, only recently \cite{bbchallenge} could show that the long known \enquote{likely champion} already is the winner. And from $n=6$ onward, the numbers explode: The current record holder, as of June 2025, see \cite{Michelwww}, runs for more than $2\uparrow^2 (2\uparrow^2 (2\uparrow^2 10))$ steps before halting. Here, we have to use Knuth's up arrow notation, see \cite{KnuthUpArrow}, to state these large numbers. For positive integers $a$, $b$, and $n$ define
\begin{align*}
a\uparrow^n b:&= \begin{cases} a^b,& \text{ if } n=1\\ a,& \text{ if } b=1\\ a\uparrow^{n-1} (a\uparrow^n (b-1)),& \text{ otherwise}  \end{cases}.\\
\text{Thus, } a\uparrow^2 b&=a^{a^{\iddots^a}} \raisebox{1.4ex}{\bigg\} \text{of height $b$}}.
\end{align*} 
For example, $2\uparrow^2 4=2^{2^{2^2}}=2^{16}=65\,536$, $2\uparrow^2 5=2^{2\uparrow^2 4}=2^{65\,536}$, and so on. Thus, the current record holder for the Busy Beaver competition with 6~states runs for more than
\[2^{2^{\iddots^2}} \raisebox{1.4ex}{\bigg\} \text{of height $2^{2^{\iddots^2}} \raisebox{1.4ex}{\bigg\} \text{of height $2^{2^{\iddots^2}} \raisebox{1.4ex}{\bigg\} \text{of height $10$}}$}}$}}\]
steps. And for $n=7$ the best known halting machine runs for over $2\uparrow^{11} (2\uparrow^{11} 3)$ steps. Further details can be found in Michel's overview article \cite{Michel}, his more up to date website, \cite{Michelwww}, and on the associated website \cite{Marxen}, which also hosts the paper by \cite{MarxenBuntrock} on their search for Busy Beavers with $n=5$ states.

We have posed a somewhat more specific problem: How does the answer to the question of the maximum number of possible steps before halting change when we additionally require that the tape has to be completely empty not only at the beginning but also after the computation is complete? (In contrast to the \emph{Busy Beavers}, which earned their nickname because the non-blank symbol is typically denoted by \verb+|+, resembling a tree trunk, thus leading to whole dams of such tree trunks being built side by side (\verb+||||||||||+) during their computations, the machines considered here are not productive: At the end, nothing remains. Or, in other words, much ado about nothing. Hence, the humorous term \enquote{civil servant beaver} – or in Latin: \emph{Castor Ministerialis}.)

\section{Search strategy: How to find beavers?}

When embarking on the search for (civil servant) beavers, two main tasks arise:
\begin{itemize}
\item Reducing the number of Turing machines to be examined.
\item Recognizing whether a given Turing machine halts or not.
\end{itemize}

There are approaches for both sub-tasks, which we discuss in the following section.

\subsection*{Reduction of the number of Turing machines to be examined}
To determine the number of different Turing machines with $n$ states, we can proceed as follows: For each of the $n$ states a machine can be in, there are two possible symbols that can be read. Thus, there are $2n$ different pairs of states and read symbols. In each of these cases, three actions must be taken: writing a new symbol to the tape, moving the RW-head to the left or to the right, and transitioning the machine to a new state. There are $2\cdot 2\cdot (n+1)$ possible variants here, assuming the halting state is not counted as one of the $n$ states. In total, we obtain $(4n+4)^{2n}$ different Turing machines with $n$ states, which grows rapidly even for small $n$. For $n=3$, this results in more than 16 million machines, for $n=4$ more than 25 billion, and for $n=5$ already over $6\cdot 10^{13}$ different Turing machines.

However, not all of these Turing machines need to be analyzed in order to find one with the maximum number of steps. In fact, many of these machines behave identically to each other, especially when, for example, the states (including references to them) are only permutated among themselves while keeping the initial state the same. Furthermore, there is a corresponding machine for each Turing machine where the directions are swapped, which behaves analogously to the original machine. Therefore, only one machine from each such pair needs to be considered.

Thus, it is useful to consider the Turing machines in a normalized order, so that not all machines are processed, but at least one from each \enquote{type} is considered. Specifically, we can proceed as follows:

\begin{itemize}
\item In the starting state, when reading the blank symbol---which we from now on denote by the digit zero---, which is always the first symbol read on the empty tape in the first cycle, the RW-head always moves to the \emph{right}. (This excludes the mirror-image machine.)
\item The machine then either moves to the halting state (for $n=1$) or to the second state (for $n>1$). (It cannot remain in the same state, as it would then enter an infinite loop and never halt. By this restriction, the arrangement of the non-starting states is somewhat constrained.)
\item Whenever a (state, symbol) pair that has not yet been reached in the simulation of the machine is encountered, the definition of the Turing machine is extended to include the corresponding assignment, where only the first $m+1$ states are available as target states. Here, $m$ is the number of states that have been used in the machine's simulation so far. (This further restricts the permutation of the set of states.)
\item If in the process two states are constructed where the same actions are taken for each read symbol, then these two states are semantically equivalent. In this case, they can be merged into one state, resulting in a Turing machine with one fewer state. However, since a civil servant beaver with $n+1$ states always takes more steps than one with $n$ states, these cases no longer need to be considered.
\item It is not clear in advance whether a civil servant beaver will write the other symbol---which we from now on will denote with the digit one---in the first cycle. However, if it does not, the second state will again encounter an empty tape and starts the computation from there, resulting in a Turing machine that halts with one step less. The same reasoning applies, if the third state used is the first to write a one, and so on.  Therefore, we can require, without loss of generality, that the machine we are investigating writes a one in the first cycle, and then check if a different starting state might first write a zero and transition to the actual starting state (or via a subset of other states, each writing zeros and applying only once).
\end{itemize}

In combination with the non-halting detection discussed in the next section, this approach significantly reduces the number of Turing machines to be considered: For $n=5$, only about 46~million were analyzed, instead of the $6\cdot 10^{13}$ mentioned earlier, which is less than one in every million.

\subsection*{(Non-)Halting detection and simulation}

As described in the previous section, the different Turing machines are generated incrementally: When a previously undefined (state, symbol) pair is encountered during the simulation of a machine, \enquote{daughter machines} are created in which the various possible actions for this pair are defined. But how can one effectively detect whether such a Turing machine (again on the empty tape) halts? There are several ideas for this, all of which can be combined:

\begin{itemize}
\item One simulates the Turing machine for a certain number of cycles. If the machine halts during this time, this information is known; if not, it remains a potential non-halting machine.
\item If only a maximum of four states have been (at least partially) defined and the Turing machine has been running for more than $107$ cycles, then it can no longer halt. This is known from the completed search for Busy Beavers with $n=4$ states. In this case, the further simulation can be terminated.
\item Since the Turing machine is supposed to halt on the empty tape, a backtracking approach is possible: What configuration must have prevailed in the last cycle, the second-to-last cycle, and so on? If a contradiction is generated in this process, the considered Turing machine is not suitable for a civil servant beaver.
\item If a Turing machine, starting from a proper subset of states that does not include the halting state, cannot exit this subset because only transitions to states within this subset are defined, it also cannot halt.
\item If the same configuration (including the same tape content and head position) is reached twice, a loop exists, meaning the machine will not halt in the future.
\item If the Turing machine's RW-head reaches a position more than half the number of cycles simulated so far away from the starting cell, the machine can only move the head infinitely in the corresponding direction and therefore, will not halt.
\end{itemize}

With this approach, while scanning the Turing machines with 5 states for civil servant beavers, it could be determined for about 89\% of the 46~million machines analyzed whether they would halt within a maximum of 47.2~million steps or not. For the remaining approximately 6.4 million machines, no such conclusion could be drawn. However, since the proof of $BB(5)=47\,176\,870$---that is, every Turing machine with 5 states that runs longer, when starting from the empty tape, could not hold anymore---we know that none of the remaining Turing machines, not characterized by our program, can hold, too. Thus, there is no \enquote{civil servant beaver} in this group and we already found a champion.

A summary of the findings from scanning the Turing machines for civil servant beavers with up to six states can be found in the next section.

\section{Results}
The results (one machine for each state count with a maximum step count) are presented below. The blank symbol is always the symbol~0, $A$ is the starting state, and the transition to the end state is given by \enquote{\textsc{halt}}:

\begin{itemize}
\item 1 state: halts after 1 step \\

\begin{tabular}{r|rrc|rrc}
 & \multicolumn{6}{c}{Read Symbol}\\
State & \multicolumn{3}{c}{0} & \multicolumn{3}{c}{1}\\
\hline
$ A$ & 0 & $r$    & \textsc{halt} & 0 & $r$  & \textsc{halt} \\
\end{tabular}

For each state (here only $A$), the two cases are indicated, depending on which symbol (0 or 1) is read. In both cases, a triple is given between the two bars, specifying which symbol should be written, whether the RW-head moves left ($\ell$) or right ($r$), and to which subsequent state the machine transitions. If \emph{\textsc{halt}} is written, the transition refers to the (implicitly defined) halting state, and the machine halts. This Turing machine halts after just one step when it reads the first symbol (a zero) since it starts in state $A$ and then transitions to the halting state.

\item 2 states: halts after 4 steps \\

\begin{tabular}{r|rrc|rrc}
 & \multicolumn{6}{c}{Read Symbol}\\
State & \multicolumn{3}{c}{0} & \multicolumn{3}{c}{1}\\
\hline
$ A$ & 0 & $r$    & $B$ & 0 & $r$  & \textsc{halt} \\
$ B$ & 1 & $\ell$ & $A$ & 0 & $r$ &  \textsc{halt} \\
\end{tabular}

This Turing machine reads a zero from the empty tape in the first cycle, starts in state~$A$, and then follows the action specified in the table: It writes a zero to the current cell, moves one cell to the right, and transitions to state~$B$. In the second cycle, it reads another zero (since the tape is initially filled with zeros), but now it is in state~$B$. Therefore, the following actions are executed: Write a one to the cell, move one cell to the left, and return to state~$A$, and so on.

\begin{minipage}{0.47\textwidth}
\item 3 states: halts after 12 steps \\

\begin{tabular}{r|rrc|rrc}
 & \multicolumn{6}{c}{Read Symbol}\\
State & \multicolumn{3}{c}{0} & \multicolumn{3}{c}{1}\\
\hline
$ A$ & 1 & $r$    & $B$ & 0 & $\ell$  & $C$ \\
$ B$ & 0 & $\ell$ & $A$ & 0 & $r$     & $B$ \\
$ C$ & 1 & $\ell$ & $A$ & 0 & $r$     & \textsc{halt} \\
 \multicolumn{7}{c}{\phantom{.}}\\
\end{tabular}
\end{minipage}
\begin{minipage}{0.47\textwidth}
\item 4 states: halts after 34 steps \\

\begin{tabular}{r|rrc|rrc}
 & \multicolumn{6}{c}{Read Symbol}\\
State & \multicolumn{3}{c}{0} & \multicolumn{3}{c}{1}\\
\hline
$ A$ & 1 & $r$    & $B$ & 1 & $\ell$  & $B$ \\
$ B$ & 1 & $\ell$ & $A$ & 0 & $\ell$  & $C$ \\
$ C$ & 0 & $r$    &\textsc{halt}& 0 & $\ell$  & $D$ \\
$ D$ & 1 & $r$    & $D$ & 0 & $\ell$  & $B$ \\
\end{tabular}
\end{minipage}
\medskip

\begin{minipage}{0.47\textwidth}
\item 5 states: halts after 187 steps \\

\begin{tabular}{r|rrc|rrc}
 & \multicolumn{6}{c}{Read Symbol}\\
State & \multicolumn{3}{c}{0} & \multicolumn{3}{c}{1}\\
\hline
$ A$ & 1 & $r$    & $B$ & 1 & $\ell$  & $E$ \\
$ B$ & 1 & $\ell$ & $C$ & 0 & $\ell$  & $B$ \\
$ C$ & 0 & $r$    & $D$ & 0 & $\ell$  & $C$ \\
$ D$ & 0 & $r$    & $A$ & 0 & $r$     & \textsc{halt} \\
$ E$ & 1 & $r$    & $A$ & 1 & $r$     & $E$ \\
 \multicolumn{7}{c}{\phantom{.}}\\
\end{tabular}
\end{minipage}
\begin{minipage}{0.47\textwidth}

\item 6 states: halts after 438\,120 steps\\

\begin{tabular}{r|rrc|rrc}
 & \multicolumn{6}{c}{Read Symbol}\\
State & \multicolumn{3}{c}{0} & \multicolumn{3}{c}{1}\\
\hline
$ A$ & 1 & $r$    & $B$ & 1 & $r$     & $A$ \\
$ B$ & 1 & $\ell$ & $B$ & 1 & $\ell$  & $C$ \\
$ C$ & 1 & $r$    & $D$ & 0 & $r$     & $E$ \\
$ D$ & 0 & $\ell$ & $E$ & 0 & $r$     & $A$ \\
$ E$ & 0 & $r$    &\textsc{halt}& 0 & $r$     & $F$ \\
$ F$ & 0 & $r$    & $B$ & 0 & $r$     & $C$ \\
\end{tabular}
\end{minipage}

For $n \leq 5$ states, it was also proven that the given step counts represent maximal values. These results can also be found in the sequence \seqnum{A1119683} of the On-Line Encyclopedia of Integer Sequences. For $n=6$, the mentioned machine is the best we found. Thus, the given machine is a civil servant beaver, or such a machine runs for more than five million steps. And this is quite possible. 
\end{itemize}

\section{Analysis of the best candidate with 6 states}
To demonstrate how complex the behavior of such Turing machines can be even with a small number of states, we want to analyze the previously mentioned candidate for a civil servant beaver with six states.\footnote{For self-testing, one can try this Turing machine using an on-line emulator at the following URL: {\scriptsize\url{https://morphett.info/turing/turing.html?4f0042cb2598cf72b5d41e40c3cd9f15}}.}

First, when looking at the state/transition table, a few things stand out:
\begin{enumerate}
\item If the Turing machine is in state~$A$ and reads a one, the RW-head moves to the right until the first zero after the sequence of ones is found. This first zero is changed by state~$A$ into a one, the RW-head moves one cell to the right, and the machine transitions into state~$B$. Therefore, if there were $k$ ones initially directly behind the RW-head, one more is added, and this process took a total of $k+2$ steps. \label{EigenschaftA1}
\item If the machine is in state~$B$ and reads a zero, the RW-head moves to the left until the first one after this block of zeros is found. On the way, all zeros are changed to ones. This leftmost one remains, and the machine moves one more cell to the left and into state~$C$. So, if the RW-head started at the last cell of a block of $k$ zeros (provided that before this block there was at least one one), the Turing machine will transform this into a block of $k+1$ ones (including the one that was originally before the zero block) in $k+1$ steps, and the machine will now be at the cell directly before that. \label{EigenschaftB0}
\item If the machine is in state~$C$, $E$, or $F$ and reads a one, the RW-head moves to the right until the first zero is found. During this process, all ones are rewritten to zeros, and the states $C$, $E$, and $F$ are cycled through in this order. When the first zero after this now erased block of ones is found, the behavior of the machine depends on which of the three states it is currently in, i.e., the residue class modulo~3 of the number of ones in this block. So, if there was initially a block of $k$ ones following the first one, this will be transformed into a block of $k+1$ zeros (including the former first one) in $k+1$ steps, and the RW-head will then be positioned on the first zero after this block, while the machine will be in state $E$, $F$, or $C$ if $k \equiv 0$, $1$, or $2$ modulo~3, respectively. \label{EigenschaftCEF1}
\end{enumerate}

With these insights and some additional detailed considerations, we now can investigate the behavior of this Turing machine. We will focus only on the cases that are relevant when the machine starts on the empty tape.

Certain configurations in the computation of this Turing machine will play a more significant role, so we want to assign them a special designation. The configuration $C(k_0,k_1,k_2)$ will refer to the state in which there are initially (beginning with the first non-zero symbol) $k_0$ ones, followed by a zero, and then $k_1+k_2$ more ones (which are followed only by zeros), where the RW-head is positioned on the $k_1$-th one of the second block, and the machine is in state $C$. We will denote this briefly as $C(k_0,k_1,k_2):=1^{k_0}01^{k_1}\_C\_1^{k_2}$. The parameters $k_0$ to $k_2$ can also take the value 0, which simply means the corresponding block disappears. (If $k_1 = 0$, the RW-head is on the zero directly before the block of $k_2$ ones.)

Let us first consider the initial steps of this Turing machine. It starts on the empty tape in state~$A$: $\text{\textsc{start}}:=0\_A$. The zero is changed to a one, the RW-head moves to the right, and the machine transitions to state~$B$: $0\_A \Schritte{} 10\_B$. According to Property~(\ref{EigenschaftB0}), this zero is now changed to a one, the RW-head moves to the left, and finds the first one. This one remains, and the machine moves one more cell to the left and into state~$C$: $0\_C\_1^2$. This is exactly the configuration $C(0,0,2)$, so we can summarize by saying that in three steps the machine transitions from the start to the $C(0,0,2)$ configuration. We write this as

\[\text{\textsc{start}} \Schritte{3} C(0,0,2).\]

Starting from a configuration $C(k_0,k_1,k_2)$, we perform an analysis by cases. In the first level, we distinguish whether $k_1$ disappears or is positive.

\textbf{Case 1: $k_1=0$.} Now, it is relevant how large $k_2$ is. For our consideration of the start on the empty tape, we only need the subcase $k_2 \geq 2$. We then observe the following transitions:
\begin{align*}
C(k_0,0,k_2)= 1^{k_0}0\_C\_1^{k_2} & \Schritte{} 1^{k_0+2}\_D\_1^{k_2-1}  \Schritte{} 1^{k_0+1}01\_A\_1^{k_2-2}\\
                                                                 & \Schritte{k_2} 1^{k_0+1}01^{k_2}0\_B; \text{  by Property~(\ref{EigenschaftA1})}\\
						 & \Schritte{2}    1^{k_0+1}01^{k_2-1}\_C\_1^2=C(k_0+1,k_2-1,2).
\intertext{The last step follows  by Property~(\ref{EigenschaftB0}). In summary, we obtain}
C(k_0,0,k_2) &\Schritte{k_2+4} C(k_0+1,k_2-1,2).\\
\intertext{In particular, we get}
C(0,0,2) &\Schritte{6} C(1,1,2).
\end{align*}

\textbf{Case 2: $k_1\geq 1$.} Here, the Turing machine's RW-head reads a one, and it is in state $C$, so the situation corresponds to Property~(\ref{EigenschaftCEF1}). Accordingly, the congruence class of $k_2$ modulo 3 is relevant, and we distinguish the following subcases:

\textbf{Case 2.1: $k_2 \equiv 0 \pmod{3}$.} Then the following occurs:
\begin{align*}
C(k_0, k_1, k_2) = 1^{k_0}01^{k_1}\_C\_1^{k_2} &\Schritte{k_2+1} 1^{k_0}01^{k_1-1}0^{k_2+2}\_E \Schritte{} \text{\textsc{halt}}\\
\intertext{or, in summary,}
C(k_0, k_1, k_2) &\Schritte{k_2+2} \text{\textsc{halt}} \text{ for $k_1\geq 1$ and $k_2\equiv 0\pmod{3}$},\\
\intertext{where the Turing machine halts on the empty tape if additionally $k_0=0$ and $k_1=1$.}
\end{align*}

\textbf{Case 2.2: $k_2 \equiv 1 \pmod{3}$.} Then the following situation occurs:

\begin{align*}
C(k_0, k_1, k_2) = 1^{k_0}01^{k_1}\_C\_1^{k_2} &\Schritte{k_2+1} 1^{k_0}01^{k_1-1}0^{k_2+2}\_F \Schritte{} 1^{k_0}01^{k_1-1}0^{k_2+3}\_B\\
      & \Schritte{k_2+3} 1^{k_0}01^{k_1-1}\_B\_1^{k_2+3}. \tag{*} \label{Fall2.2Divergenz}
\intertext{If $k_1\geq 2$, the Turing machine's RW-head continues to read a one and moves left to state $C$:}
 &\Schritte{} 1^{k_0}01^{k_1-2}\_C\_1^{k_2+4} = C(k_0, k_1-2, k_2+4),
\intertext{or, in summary,}
C(k_0, k_1, k_2) &\Schritte{2k_2+6} C(k_0, k_1-2, k_2+4) \text{ for $k_1\geq 2$ and $k_2\equiv 1\pmod{3}$}.\\
\intertext{If $k_1 = 1$, the block $1^{k_1-1}$ is empty, and the RW-head reads a zero in configuration~\eqref{Fall2.2Divergenz}. We proceed with}
\dots &\Schritte{} 1^{k_0}\_B\_1^{k_2+4} \Schritte{} 1^{k_0-1}\_C\_1^{k_2+5} = C(0,k_0-1, k_2+5)\\
\intertext{or, in summary,}
C(k_0, 1, k_2) &\Schritte{2k_2+7} C(0, k_0-1, k_2+5) \text{ for $k_0\geq 1$ and $k_2\equiv 1\pmod{3}$}.\\
\intertext{Note, that here we assume $k_0 \geq 1$. We do not consider the subcase $k_0 = 0$ as it does not play a role in our investigation of a run of the Turing machine starting from the empty tape.}
\end{align*}

\textbf{Case 2.3: $k_2 \equiv 2 \pmod{3}$.} Similarly, we get:
\begin{align*}
C(k_0, k_1, k_2) \!=\! 1^{k_0}01^{k_1}\_C\_1^{k_2} &\Schritte{k_2+1} 1^{k_0}01^{k_1-1}0^{k_2+2}\_C \Schritte{} 1^{k_0}01^{k_1-1}0^{k_2+1}10\_D\\
 & \Schritte{} 1^{k_0}01^{k_1-1}0^{k_2+1}1\_E \Schritte{} 1^{k_0}01^{k_1-1}0^{k_2+3}\_F\\ 
 & \Schritte{} 1^{k_0}01^{k_1-1}0^{k_2+4}\_B \Schritte{k_2+4} 1^{k_0}01^{k_1-1}\_B\_1^{k_2+4}.\tag{**}\label{Fall2.3Divergenz}
\intertext{As in case 2.2, we distinguish between $k_1 \geq 2$ and $k_1 = 1$. If $k_1 \geq 2$, the calculation continues with}
 & \Schritte{} 1^{k_0}01^{k_1-2}\_C\_1^{k_2+5} = C(k_0,k_1-2,k_2+5)
\intertext{or, in summary,}
C(k_0, k_1, k_2) &\Schritte{2k_2+10} C(k_0,k_1-2,k_2+5) \text{ for $k_1\geq2$ and $k_2\equiv 2\pmod{3}$}.
\intertext{If $k_1 = 1$, we proceed from \eqref{Fall2.3Divergenz}:}
& \Schritte{} 1^{k_0}\_B\_1^{k_2+5} \Schritte{} 1^{k_0-1}\_C\_1^{k_2+6}= C(0,k_0-1,k_2+6)
\intertext{or, in summary,}
C(k_0, 1, k_2) &\Schritte{2k_2+11} C(0,k_0-1,k_2+6)
\end{align*}
for $k_0\geq 1$ and $k_2\equiv 2\pmod{3}$. As above, we assume $k_0\geq 1$, since $k_0=0$ is not relevant for our investigation.

What is striking in Cases 2.2 and 2.3  is that if $k_1$ is sufficiently large, it is reduced by two in each of these summarized \enquote{steps}, while the congruence of $k-2$ modulo~3 always jumps back and forth between the two cases: If $k_2\equiv 1\pmod{3}$, we get $k_2+4\equiv 2\pmod{3}$ as the next value of this parameter. And if $k_2\equiv 2\pmod{3}$, we now get $k_2+5\equiv 1\pmod{3}$.  In addition to the congruence of $k_2$ modulo~3, the congruence of~$k_1$ modulo~4 is also relevant: If $k_1$ is at least~2, it is reduced by~2 in each such \enquote{step}, so that the parity remains unchanged. And the jumping back and forth between the two cases explains the additional factor of~2. Only in the case of $k_1=1$ or $k_1=0$ do other effects occur.

Thus, we obtain for $k_2\equiv 2\pmod{3}$ (with the additional assumption $k_2\geq 2$ and $k_0\geq 1$, if $k_1=1$)
\begin{align*}
C(k_0, 0, k_2) &\Schritte{ k_2+4}   C(k_0+1, k_2-1,     2),\\
C(k_0, 1, k_2) &\Schritte{2k_2+11}  C(    0, k_0-1, k_2+6),\\
C(k_0, 2, k_2)&\Schritte{2k_2+10}  C(k_0  ,     0, k_2+5) \Schritte{k_2+9} C(k_0+1, k_2+4, 2) \text{, thus}\\
C(k_0, 2, k_2)&\Schritte{3k_2+19} C(k_0+1, k_2+4, 2),\\
C(k_0, 3, k_2) &\Schritte{2k_2+10}  C(k_0  ,     1, k_2+5)\Schritte{2k_2+17} C(  0,k_0-1,k_2+10)\text{, hence}\\
C(k_0, 3, k_2) &\Schritte{4k_2+27} C(  0,k_0-1,k_2+10),\\ 
\intertext{and for $k_1\geq 4$}
C(k_0, k_1, k_2) &\Schritte{2k_2+10} C(k_0,k_1-2,k_2+5) \Schritte{2k_2+16} C(k_0,k_1-4,k_2+9)\text{, and}\\
C(k_0, k_1, k_2) &\Schritte{4k_2+26} C(k_0,k_1-4,k_2+9).\\
\intertext{Inductively, for $k_1=4m, \dots, k_1=4m+3$ we now get}
C(k_0, 4m, k_2) &\Schritte{4mk_2+18m^2+8m} C(k_0, 0, k_2+9m) \Schritte{k_2+9m+4} C(k_0+1, k_2+9m-1, 2),\\
C(k_0, 4m+1, k_2) &\Schritte{4mk_2+18m^2+8m} C(k_0, 1, k_2+9m) \Schritte{2k_2+18m+11} C(0, k_0-1, k_2+9m+6),\\
C(k_0, 4m+2, k_2) &\Schritte{4mk_2+18m^2+8m} C(k_0, 2, k_2+9m) \Schritte{3k_2+27m+19} C(k_0+1, k_2+9m+4,2),\\
C(k_0, 4m+3, k_2) &\Schritte{4mk_2+18m^2+8m} C(k_0, 3, k_2+9m) \Schritte{4k_2+36m+27} C(0, k_0-1, k_2+9m+10)\\
\intertext{or, in summary,}
C(k_0, 4m, k_2) &\Schritte{4mk_2+k_2+18m^2+17m+4}  C(k_0+1, k_2+9m-1, 2),\\
C(k_0, 4m+1, k_2) &\Schritte{4mk_2+2k_2+18m^2+26m+11} C(0, k_0-1, k_2+9m+6),\\
C(k_0, 4m+2, k_2) &\Schritte{4mk_2+3k_2+18m^2+35m+19} C(k_0+1, k_2+9m+4,2),\\
C(k_0, 4m+3, k_2) &\Schritte{4mk_2+4k_2+18m^2+44m+27}  C(0, k_0-1, k_2+9m+10).
\end{align*}
In a similar way, one could also obtain a list for $k_2\equiv 1\pmod{3}$. We omit this here as it does not play a role in our analysis. 

With the results, we can summarize further and obtain
\begin{alignat*}{8}
\text{\textsc{start}} &\Schritte{3}& C(0,0,2)  &\Schritte{6}& C(1,1,2)             &\Schritte{15}& C(0,0,8)     &\Schritte{12}& C(1,7,2) \\
&\Schritte{105}& C(0,0,21)       &\Schritte{25}& C(1,20,2)  &\Schritte{581}&  C(2,46, 2)             &\Schritte{2\,676}& C(3, 105,2)\\
&\Schritte{13\,067}& C(0, 2, 242) &\Schritte{745}& C(1, 246, 2)     &\Schritte{69\,626}& C(2, 555,2)  &\Schritte{350\,003}&  C(0, 1, 1\,254)&\Schritte{1\,256}& \text{\textsc{halt}},\\
\end{alignat*}
or in short
\[\text{\textsc{start}} \Schritte{438\,120} \text{\textsc{halt}},\]
where the Turing machine we analyzed stops on the empty tape due to the parameter assignment $k_0=0$ and $k_1=1$ in the last section.

\section{On Turing machines with larger alphabet}
As in \cite{Michelwww} we can extend our search to find such \enquote{civil servant machines} using a larger alphabet. In the previous sections the Turing machines we looked at only could use the blank symbol and one other, thus had an alphabet of size~2. Adding more symbols increases the number of possible computational paths: the number of configurations consisting of state and read symbol increases. Our program and algorithms are easily generalizable to include a larger alphabet size. As the search space increases more than exponentially in the number of possible configurations, which is the product of the number of states and the size of the alphabet used by the Turing machine, additionally to the ones of the above sections we only looked at the cases with at most ten possible configurations. Our findings are summarized in Table~\ref{Tab:BBlargeAlphabet}. The findings suggest that adding more symbols boost the capability of Turing machines more than adding the same amount of states.

 \begin{table}[H]
 \begin{center}
 \begin{tabular}{cr|rrrrrr}
&  & \multicolumn{6}{c}{states}\\
&  & 1 & 2 & 3 & 4 & 5 & 6\\
\hline
 \multirow{4}{*}{symbols} &
 2 & \textbf{1} & \textbf{4} & \textbf{12} & \textbf{34} & \textbf{187} & 438\,120\\
& 3 & \textbf{1} & \textbf{13} & 102\\
& 4 & \textbf{1} & \textbf{39} & 26\,768\\
& 5 & \textbf{1} & 504
 \end{tabular}
 \caption{Maximum number of steps a Turing machine with $m$ states and $n$ symbols can run, starting from and halting on the empty tape\\
 proven values in \textbf{bold}, others are best known candidates\\
 entries for $(m,n)\in\{(3,3),\ (2,5)\}$ are tested up to ten million steps}
 \label{Tab:BBlargeAlphabet}
 \end{center}
 \end{table}

\phantomsection
\markright{Bibliography}
\bibliographystyle{alpha}
\bibliography{Literatur}

\newcommand{\etalchar}[1]{$^{#1}$}
\begin{thebibliography}{bCBB{\etalchar{+}}25}

\bibitem[AH28]{AckermannHilbert}
W.~Ackermann and D.~Hilbert.
\newblock {\em {Grundz\"{u}ge} der Theoretischen Logik}.
\newblock Springer, Berlin, 1928.

\bibitem[bCBB{\etalchar{+}}25]{bbchallenge}
The bbchallenge Collaboration, Justin Blanchard, Daniel Briggs, Konrad Deka,
  Nathan Fenner, Yannick Forster, Georgi Georgiev, Matthew~L. House, Rachel
  Hunter, Iijil, Maja {K\k{a}dzio\l{}ka}, Pavel Kropitz, Shawn Ligocki, mxdys,
  Mateusz {Na\'{s}ciszewski}, savask, Tristan {St\'{e}rin}, Chris Xu, Jason
  Yuen, and {Th\'{e}o} Zimmermann.
\newblock Determination of the fifth {Busy} {Beaver} value, 2025.

\bibitem[Hil00]{Hilbert}
D.~Hilbert.
\newblock {Mathematische} {Probleme}. {Vortrag}, gehalten auf dem
  internationalen {Mathematiker-Kongre\ss{}} zu {Paris} 1900.
\newblock {\em Nachrichten von der Gesellschaft der Wissenschaften zu
  {G\"{o}ttingen}, Mathematisch-Physikalische Klasse aus dem Jahre 1900}, pages
  253--297, 1900.

\bibitem[Hil02]{Hilbert2}
D.~Hilbert.
\newblock Mathematical problems.
\newblock {\em Bulletin of the American Mathematical Society}, 8:437--479,
  1902.

\bibitem[Knu73]{KnuthTAOCP1}
Donald~E. Knuth.
\newblock {\em The Art of Computer Programming, Volume 1: Fundamental
  Algorithms}.
\newblock Addison-Wesley, 1973.

\bibitem[Knu76]{KnuthUpArrow}
D.~E. Knuth.
\newblock Mathematics and {Computer} {Science}: Coping with finiteness.
\newblock {\em Science}, 194(4271):1235--1242, 1976.

\bibitem[Mar16]{Marxen}
H.~Marxen.
\newblock {Busy Beaver}.
\newblock \url{https://turbotm.de/~heiner/BB}, 2016.

\bibitem[MB90]{MarxenBuntrock}
H.~Marxen and J.~Buntrock.
\newblock Attacking the {Busy Beaver} 5.
\newblock {\em Bulletin of the EATCS}, 40:247--251, 1990.
\newblock \url{https://turbotm.de/~heiner/BB/mabu90.html}.

\bibitem[Mic22]{Michel}
P.~Michel.
\newblock The {Busy Beaver Competition}: a historical survey.
\newblock \url{https://arxiv.org/pdf/0906.3749.pdf}, 2022.

\bibitem[Mic25]{Michelwww}
Pascal Michel.
\newblock Historical survey of {Busy} {Beavers}.
\newblock \url{https://bbchallenge.org/~pascal.michel/ha.html}, 2025.

\bibitem[{Rad}62]{Rado}
T.~{Rad\'o}.
\newblock On non-computable functions.
\newblock {\em Bell System Technical Journal}, 41, 1962.
\newblock
  \url{https://computation4cognitivescientists.weebly.com/uploads/6/2/8/3/6283774/rado-on_non-computable_functions.pdf}.

\bibitem[{Sch}09]{Schoning09}
Uwe {Sch\"{o}ning}.
\newblock {\em {Theoretische Informatik -- kurz gefasst}}.
\newblock Spektrum A\-ka\-de\-mi\-scher Verlag, 2009.

\bibitem[Tur37]{Turing}
A.~M. Turing.
\newblock On computable numbers, with an application to the
  {Entscheidungsproblem}.
\newblock {\em Proceedings of the London Mathematical Society},
  s2-42(1):230--265, 1937.

\end{thebibliography}

\end{document}